\documentclass[
superscriptaddress,
groupedaddress,
prereprint,
preprintnumbers,
nofootinbib,
amsmath,amssymb,
aps,
prd,
floatfix,
]{revtex4-2}
\usepackage[utf8]{inputenc}
\usepackage{graphicx}
\usepackage{dcolumn}
\usepackage{bm}
\usepackage{hyperref}
\hypersetup{
    colorlinks=true,
    allcolors=blue,
}

\newcommand{\eq}[1]{Eq.~\eqref{#1}}

\usepackage{xcolor}

\begin{document}

\title{Hamiltonian Constraints on Spontaneous Lorentz Symmetry Breaking in the Bumblebee Model}
\author{Jie Zhu}
 \email{jiezhu@cqu.edu.cn}
 \affiliation{Department of Physics and Chongqing Key Laboratory for Strongly Coupled Physics, Chongqing University, Chongqing 401331, P.R. China}

\author{Hao Li}
  \email{Corresponding author: haolee@cqu.edu.cn}
   \affiliation{Department of Physics and Chongqing Key Laboratory for Strongly Coupled Physics, Chongqing University, Chongqing 401331, P.R. China}

\author{Zhi Xiao${}^{a, b}$}
\email{spacecraft@pku.edu.cn}
\affiliation{${}^{a}$North China Electric Power University, Beijing 102206, China}
\affiliation{${}^{b}$Hebei Key Laboratory of Physics and Energy Technology, North China Electric Power University, Baoding 071000, China}

\date{\today}

\begin{abstract}

This study demonstrates that the common practice of determining spontaneous Lorentz violation via the minimum of a Lagrangian potential is generally incorrect. By analyzing the Hamiltonian structure and constraints of vector fields, we show that the true vacuum must be derived from the Hamiltonian density. We prove that the standard quadratic potential cannot consistently generate a vacuum expectation value (VEV), identifying a cubic potential as the simplest viable alternative. Furthermore, we prove that smooth potentials only support stable timelike or lightlike VEVs. These conclusions extend to higher-rank tensor fields and impose rigorous consistency constraints on higher-rank tensor fields and Lorentz-violating effective field theories.
\end{abstract}

\maketitle

Spontaneous breaking of Lorentz symmetry~\cite{SBLV1989a,SBLV1989b} plays a central role in many approaches to Lorentz-violating effective field theories. 
In the framework of the Standard-Model Extension (SME)~\cite{Colladay:1998fq}, the Lorentz-violating coefficients are often interpreted as vacuum expectation values (VEVs) of underlying tensor fields arising from spontaneous symmetry breaking (SSB) in a more fundamental theory. 
Among the simplest realizations of this idea is the bumblebee model~\cite{Gravity04, Bumblebee05, Bumblebee08}, in which a vector field acquires a nonzero VEV through a potential that fixes the norm of the field. 
Such models have been widely used as effective descriptions of dynamical Lorentz violation and have been studied in numerous contexts~\cite{Mattingly:2005re,Tasson:2014dfa}.

In most existing treatments, the potential inducing spontaneous Lorentz symmetry breaking is chosen to resemble the Higgs-type potential, 
{\it i.e.}, the quartic ``Mexican-hat” form $(\phi^2-v^2)^2$ potential. 
Correspondingly, it is taken to be a quadratic function of the invariant quantity 
$X = B^\mu B_\mu + s\,b^2$, 
so that the VEV is expected at the minimum of the potential~\cite{Gravity04, Bumblebee05}. 
However, existing literature shows that, for such quadratic potentials, the Hamiltonian of the theory is not bounded from below~\cite{Bluhm:2008yt, Carroll:2009em, Bailey:2025oun}. 
At first sight, this result appears quite puzzling, since quadratic potentials are commonly regarded as the simplest mechanism for realizing spontaneous symmetry breaking.


To make the discussion concrete, we consider the bumblebee model with action given by~\cite{Gravity04}
\begin{equation}
S=\int d^4x\sqrt{-g}\left(\frac{1}{2\kappa}\left(R+ \xi_1 B^\mu B^\nu R_{\mu\nu}+\xi_2 B_\mu B^\mu R \right)-\frac{1}{4}B_{\mu\nu}B^{\mu\nu}-V\right)+S_\mathrm{m},  \label{eq:action}
\end{equation}
where $g$ is the determinant of $g_{\mu\nu}$, $\kappa\equiv8\pi G$, $S_{\mathrm{m}}$ is the matter action (irrelevant here), $B_\mu$ the bumblebee field, and $B_{\mu\nu}\equiv\partial_{\mu}B_{\nu}-\partial_{\nu}B_{\mu}$.
The potential $V$ is selected to provide a non-vanishing VEV for $B_\mu$, and could have the following general functional form
\begin{equation*}
    V\equiv V(X), \quad X \equiv B^\mu B_\mu + s\,b^2,\label{potential}
\end{equation*}
where $s = 1, 0, -1$ characterizes the nature of the VEV as being timelike, lightlike, or spacelike, respectively~\cite{SLVScenario02}.
In previous studies, the first form of $V$ was a sigma-model potential with a Lagrange multiplier as $V(X)=\lambda\, X$~\cite{Bumblebee05},
where the quantity $\lambda$ is a Lagrange-multiplier field.
The other form is a polynomial in $X$~\cite{Bumblebee05}, where the expectation value of $B_\mu$ is assumed to be attained at the minimum of $V(X)$, i.e., at $X=0$. A typical example is $V(X)=\frac{1}{2}\lambda X^2$~\cite{Bumblebee05}, where the quantity $\lambda$ is now a constant.
However, in Ref.~\cite{Bailey:2025oun}, Q.G. Bailey {\it et al.} found that for a potential of an even-power form, $V(X)=\lambda X^{2n}$ (where $n$ is an integer), the Hamiltonian is not bounded from below.
This seemingly counterintuitive result raises the question of whether a quadratic potential can consistently trigger spontaneous symmetry breaking for a vector field.

The origin of this issue can be traced to a common misconception regarding the mechanism of SSB for vector fields. 
This misunderstanding arises from a conceptual inertia: being so accustomed to the SSB of scalar fields~\cite{Englert:1964et, Goldstone:1961eq} that 
one instinctively—and perhaps incorrectly—transplants the corresponding conclusions to the vector field case. 
In the scalar case, the standard conclusion is that the field $\phi$ aquires a vacuum expectation value at the minimum of the potential $V(\phi)$. 
It is therefore tempting to assume that a vector field likewise acquires its VEV at the minimum of $V(X)$. 
In the following, we argue that this analogy can be misleading.

Let us review the definition of SSB. 
SSB is defined as the phenomenon where the vacuum state of a theory exhibits a lower degree of symmetry than the underlying action.
The vacuum is defined as the eigenstate of the Hamiltonian with the lowest energy eigenvalue~\cite{Weinberg:1996kr}.
Within the classical limit, the vacuum configuration is identified as the global minimum of the classical Hamiltonian~\cite{Jackiw:1974cv}.
Combining the definitions above, the conclusion is clear: for a system with the Hamiltonian $H=T+H_V$ (where $T$ represents the kinetic energy and $H_V$ the potential energy), SSB occurs at the minimum of $H_V$. 
In the case of a scalar field, we have $H_V = V$, meaning SSB is indeed attained at the minimum of $V$. 
However, for a vector field, 
owing to {\it the inherent constraint structure of vector theories, the potential part of the Hamiltonian, $H_V$, does not coincide with the functional form of the potential}~\cite{Dirac:1950pj},
and thus the situation is fundamentally different.
Therefore, the true vacuum configuration of a vector theory must be determined from the Hamiltonian density rather than directly from the Lagrangian potential.

To illustrate this point explicitly, we consider the theory in flat spacetime.
Under the $(- + + +)$ metric signature, the action reduces to
\begin{equation}
S=\int d^4x\left(-\frac{1}{4}B_{\mu\nu}B^{\mu\nu}-V(X)\right).
\end{equation}
The on-shell Hamiltonian is provided in Ref.~\cite{Bailey:2025oun} as
\begin{equation}
H=\int d^3x\left(\frac{1}{2}\vec{\Pi}^2+\frac{1}{4}B_{ij}B^{ij}+V(X)+2B_0^2V^{\prime}(X)\right),
\end{equation}
where $(B_0, B_i)$ are the four field components, and $\Pi_i = \delta \mathcal{L}/\delta \dot{B}_i$ are the corresponding conjugate momenta.
Unlike the scalar case, even imposing vanishing field strength $B_{\mu\nu}=0$, the static vacuum Hamiltonian density 
\begin{equation}
\mathcal{H}_V(B_0,\vec{B})=V(X)+2B_0^2V^{\prime}(X)\label{eq:H}
\end{equation}
still contains an extra term $2B_0^2V^{\prime}(X)$ aside from $V(X)$.
The true vacuum is obtained by minimizing the Hamiltonian density $\mathcal{H}_V$ as a function of $B_\mu$:
\begin{eqnarray}\label{MiniV}
\partial_{B_0}\mathcal{H}_V=2B_0[V'(X)-2B_0^2 V''(X)]=0
\qquad
\partial_{B_i}\mathcal{H}_V=2B_i[V'(X)+2B_0^2 V''(X)]=0.
\end{eqnarray}
Since both $B_0$ and $X\equiv -B_0^2+\vec{B}^2+s\,b^2$ range from $-\infty$ to $+\infty$ 
and $B_\mu$ are not necessarily zero in a general frame, the two equations in (\ref{MiniV})
indicate that the extrema of $\mathcal{H}_V$ are attained when the following conditions are satisfied
\begin{eqnarray}\label{VacuumCond}
V'(X)=0,\qquad V''(X)=0.
\end{eqnarray}
If the vector field is to acquire a vacuum expectation value at $X=0$, then $\mathcal{H}_V$ must attain its global minimum at $X=0$,
so we would have
\begin{equation}
V'(0)=V''(0)=0.
\end{equation}
Furthermore, the extremum is indeed a local minimum 
only if $V^{(3)}(0)\geq 0$ (see appendix \ref{Hessian}).
This result immediately shows that the commonly assumed quadratic potential $V(X)=\frac{1}{2}\lambda X^2$
does not satisfy the condition required for SSB.

Given the requirement that $V'(0)=V''(0)=0$, 
the simplest potential satisfying the above conditions is a cubic potential $V(X) = \frac{\lambda}{3} X^3$ with $\lambda>0$.
It then leads to 
\begin{equation}
\mathcal{H}_V = \frac{\lambda}{3}X^2\left(X+6 B_0^2\right)=\frac{\lambda}{3}X^2 (5 B_0^2 + \vec{B}^2+s\, b^2).
\end{equation}
If $s=-1$, let $B_0 = x\, b $ and $|\vec{B}|=y\, b$, then
\begin{equation}
\mathcal{H}_V = \frac{\lambda\, b^6}{3} (-x^2+y^2-1)^2(5x^2+y^2-1).
\end{equation}
The above $\mathcal{H}_V$ obtains its minimum at $x=y=0$, which means that $B_\mu\equiv 0$ is the true vacuum and the SSB is not triggered.
However, if $s=1$ or $s = 0$, since $5 B_0^2 + \vec{B}^2+s\, b^2\geq 0$, $\mathcal{H}_V$ indeed attains a global minimum at $X=0$.
Consequently, a cubic potential can indeed trigger SSB in vector fields, and the resulting vacuum expectation value must be either timelike or lightlike. 

We now consider the most general smooth potential satisfying
$V'(0)=V''(0)=0$.
A general smooth potential satisfying the above conditions can be parameterized in the form $V(X)=X^3 f(X)$, where $f(X)$ is a smooth function.
The potential in the Hamiltonian is now
\begin{equation}\label{HVgeneral}
\mathcal{H}_V = X^2 \left[(6 B_0^2+X) f(X)+ 2B_0^2 Xf'(X)\right]\equiv X^2 F(B_0,|\vec{B}|).
\end{equation}
If $F(B_0, |\vec{B}|)$ can take negative values, then $\mathcal{H}_V$ could likewise drop below zero. 
However, since $\left.\mathcal{H}_V\right|_{X=0}=0$, 
{in general} $X=0$ is neither the global minimum of the Hamiltonian nor the true physical vacuum. Consequently, to ensure that $\mathcal{H}_V$ attains its global minimum at $X=0$ for all possible field configurations, we must impose the following condition
\begin{equation}
F(B_0,|\vec{B}|)
= (5 B_0^2 + \vec{B}^2+s\,b^2)f(X)+ 2B_0^2 Xf'(X)\geq 0.\label{eq:geq}
\end{equation}
We have already demonstrated that for a constant function $f(X)\equiv\lambda/3$, the condition $s \geq 0$ must hold.
In what follows, we shall prove that for a non-constant analytic function $f(X)$, the constraint \eq{eq:geq} necessarily implies $x_0\equiv s\,b^2 \geq 0$.
Substitute $B_0=B_i=0$ into \eq{eq:geq}, we have $x_0 f(x_0)\geq 0$.
On the other hand, since the Hamiltonian \eqref{eq:H} is bounded from below for all $B_0$ and $X$, it follows that for any $X \in \mathbb{R}$, the condition $V'(X) \geq 0$ must hold.
Consequently, $V(X)$ is a non-decreasing function on $\mathbb{R}$.
If $s<0$, we have $x_0\equiv s\,b^2<0$, and $V(x_0)=x_0^2\cdot x_0f(x_0)\geq 0$.
However, since $V(X)=X^3 f(X)$, we have $V(0)=0$, thus $V(x_0)\geq V(0)$.
Combining this with the fact that $V(X)$ is a non-decreasing function, we must have $V(X)$ being constant on the interval $[x_0, 0]$.
Since $f(X)$ is analytic and $V(X)=X^3 f(X)$, we have $V(X)\equiv0$ and $f(X)\equiv 0$.
However, this contradicts the assumption that $f(X)$ is a non-constant function; therefore, we must have $s \geq 0$.

One may wonder whether this conclusion could be avoided by allowing non-analytic or piecewise potentials that vanishes identically over the interval $[s\,b^2, 0]$ for $s = -1$.
In this scenario, $f(X)$, and consequently the Hamiltonian density $\mathcal{H}_V$, would be identically zero throughout this interval.
If we assume that such a potential triggers SSB with a spacelike VEV, then $\mathcal{H}_V$ must achieve its global minimum at $X = 0$.
However, it follows that $\mathcal{H}_V$ would also remain at this minimum for any value within the entire interval $[s\,b^2, 0]$. 
Specifically, $X = s\,b^2$ (corresponding to $B_\mu \equiv 0$) would also be a true vacuum state. 
In other words, the vacuum state $B_\mu = 0$ is not excluded or rendered unstable, meaning the SSB is not effectively triggered.

We therefore arrive at the following conclusions:
\begin{enumerate}
\item In scenarios where the SSB of a vector field is triggered by the potential rather than the Lagrangian multipliers, 
the VEV must be either timelike or lightlike. 

\item The general form of the potential that triggers SSB for a vector field satisfies $V'(0)=V''(0)=0$ and $V^{(3)}(0)\geq 0$.
The simplest realization of such a potential is the cubic form $V(X) = \frac{\lambda}{3} X^3$ with $\lambda > 0$.
If $f(X) \geq 0$ and $X f'(X) \geq 0$, the potential $V(X)\equiv X^3 f(X)$ can trigger the SSB.
\end{enumerate}

These results have important implications for the physical interpretation
of Lorentz-violating effective field theories such as the SME.
In the general framework of the SME, the Lorentz-violating tensor coefficients are typically interpreted as VEVs arising from the SSB of underlying tensor fields in a more fundamental theory. 
Our analysis demonstrates that even for the simplest vector case, the dynamical conditions for triggering SSB are far more restrictive than previously assumed; specifically, the common practice of employing quadratic potentials to generate VEVs is rendered physically problematic due to the unboundedness of the Hamiltonian. 
This suggests that {the background fields in the SME cannot be treated as mere kinematic constants} or simple analogues of the Higgs field. 
As one moves from vector fields to higher-rank tensor fields, the constraint structures and the resulting Hamiltonian densities become significantly more complex. 
Therefore, our results serve as a cautionary note: the construction of UV-complete models for the SME requires a much more rigorous treatment of the potential forms and the associated stability conditions. 

The above analysis can be generalized to higher-rank tensor fields.
A generalized example is the SSB of $p$-form fields.
For a totally antisymmetric $p$-form field $A_{\mu_1 \mu_2 \dots \mu_p}$, the action in the flat spacetime is
\begin{equation}
S = \int d^4x \left( -\frac{1}{2(p+1)!} F_{\mu_1 \dots \mu_{p+1}} F^{\mu_1 \dots \mu_{p+1}} - V(X) \right),\label{eq:p_form_action} 
\end{equation}
where $F = dA$ is the field strength
\begin{equation}
F_{\mu_1 \mu_2 \dots \mu_{p+1}} = (p+1) \partial_{[\mu_1} A_{\mu_2 \dots \mu_{p+1}]},\label{eq:Fdef}
\end{equation}
and the scalar $X$ is
\begin{equation}
X = A_{\mu_1 \dots \mu_p} A^{\mu_1 \dots \mu_p}+x_0 = -p (A_{0 i_2 \dots i_p})^2 + (A_{i_1 \dots i_p})^2 +x_0,
\end{equation}
where $x_0$ is a constant and the $p$-form field $A$ acquire a VEV at $X=A_{\mu_1 \dots \mu_p} A^{\mu_1 \dots \mu_p}+x_0=0$.
The on-shell Hamiltonian is given as
\begin{equation}
H = \int d^3x \left( \frac{p!}{2} \Pi_{i_1 \dots i_p} \Pi^{i_1 \dots i_p} + \frac{1}{2(p+1)!} F_{i_1 \dots i_{p+1}} F^{i_1 \dots i_{p+1}} + V(X) + 2p V'(X) (A_{0 i_2 \dots i_p})^2 \right),
\end{equation}
where
\begin{equation}
\Pi^{i_1 \dots i_p} = \frac{\partial \mathcal{L}}{\partial \dot{A}_{i_1 \dots i_p}} = \frac{1}{p!} F^{0 i_1 \dots i_p}
\end{equation}
are the corresponding conjugate momenta.
The remainder of the derivation follows almost identically to the Bumblebee case, with the substitutions $B_0^2 \to p (A_{0 i_2 \dots i_p})^2$, $\vec{B}^2 \to (A_{i_1 \dots i_p})^2$, and $s\,b^2 \to x_0$.
Consequently, for the SSB of $p$-form fields, the conclusions remain analogous to those obtained in the Bumblebee case: the potential $V(X)$ must be at least cubic in $X$ and $x_0\geq 0$.
Furthermore, we would like to emphasize that a nonlinear potential can induce nonlinear constraint equations for non-dynamical field components
such as $B^0$ in vector theories and $A_{0i_2...i_p}$ in p-form theories. 
Consequently, even if the constraints are preserved under time evolution, the uniqueness of the evolution is not automatically guaranteed in either bumblebee
or p-form models \cite{YBonder2015E}. 
Therefore constraint-induced modifications of the Hamiltonian, and the subtleties they entail, may be a generic feature of tensor field theories
with spontaneous symmetry breaking.

{\it Conclusion:}
In this work we have shown that the mechanism of spontaneous symmetry breaking for vector fields must be analyzed at the Hamiltonian level.
Because the temporal components of vector fields generate constraints, the potential term entering the Hamiltonian density differs from the Lagrangian potential, and the physical vacuum cannot in general be identified with the minimum of the Lagrangian potential.
As a consequence, the commonly assumed quadratic potential cannot consistently generate a vacuum expectation value.
The simplest viable realization is a cubic potential, and more generally smooth potentials can only produce timelike or lightlike vacua.
We further showed that the same constraint-induced structure persists for higher-rank antisymmetric tensor fields, implying nontrivial consistency conditions for models of spontaneous Lorentz violation.
These results highlight the essential role of constraint dynamics in determining the vacuum structure of tensor field theories,  
and the vacuum structure of constrained field theories is fundamentally a Hamiltonian concept.
More broadly, they illustrate how the Hamiltonian formulation can reveal physical consistency conditions that are not manifest in the Lagrangian description, echoing Dirac's classic insight that the relation between the two formulations may still contain unexplored structure.

\section*{Acknowledgements}

This work was supported in part by the National Natural Science Foundation of China under Grant No.~12547101. HL was also supported by the start-up fund of Chongqing University under No.~0233005203009, and JZ was supported by the start-up fund of Chongqing University under No.~0233005203006.

\appendix

\section{The Hessian and null directions}\label{Hessian}
Suppose the local stationary points of the vacuum Hamiltonian density (\ref{eq:H}) occur at $X=X_*$, 
where $V'(X_*)=V''(X_*)=0$.
The Hessian of $\mathcal{H}_V$ is
\begin{equation}
H_{e}=2
\left(
\begin{array}{cc}
V'-8B_0^2V''+4B_0^4V''' & 2B_0B_i(V''-2B_0^2V''') \\
2B_0B_j(V''-2B_0^2V''') & \delta_{ij}(V'+2B_0^2V'')+2B_iB_j(V''+2B_0^2V''') \\
\end{array}
\right).
\end{equation}
Evaluating it at $X=X_*$ yields 
\begin{equation}
H_{e}|_{X=X_*}=
8\chi_0^2V'''_*\left(
\begin{array}{cc}
\chi_0^2   & -\chi_0\chi_i \\
-\chi_0\chi_i & \chi_i\chi_j \\
\end{array}
\right)=8\chi_0^2V'''_*\bar{B}_\mu\bar{B}_\nu
\Rightarrow \mathrm{det}[H_{e}]_{X=X_*}=0.
\end{equation}
where $\bar{B}_\mu\equiv(B_\mu)|_{X_*}=(\chi_0,-\chi_i)$ and $V'''_*\equiv V'''(X_*)$.
Thus, whether the stationary points $X_*=\vec{\chi}^2-{\chi_0}^2+s\,b^2$ (form a submanifold) are local minima or maxima remains undetermined.
Expanding $\mathcal{H}_V(B_0,\vec{B})$ around $\bar{B}_\mu$ by setting $B_\mu=\bar{B}_\mu+\chi_\mu$, the potential becomes
\begin{eqnarray}&&\label{Expanto3}
\mathcal{H}_V(B_0,\vec{B})
=\mathcal{H}_V(\bar{B}_\mu)+4\chi_0^2V^{(3)}_* (\bar{B}\cdot\chi)^2+\mathcal{O}(\chi_\mu^3),
\end{eqnarray}
where $\bar{B}\cdot\chi=\bar{B}_\mu\,\chi^\mu$. Hence, a local minimum requires $V^{(3)}_*=V'''(X_*)>0$.
As in the toy model of spontaneous symmetry breaking for a complex scalar field ({\it e.g.}, $V(\phi)=\frac{\lambda}{4!}(|\phi|^2-v^2)^2$),
there exists a flat direction orthogonal to the local minimum
\begin{eqnarray}&&\label{FlatS}
\chi_\mu \bar{B}^\mu=0,
\end{eqnarray}
analogous to the phase direction $\varphi$ in $\phi=|\phi|e^{i\varphi}$. 
The corresponding Goldstone mode is precisely the excitation orthogonal to  $\bar{B}^\mu$.
Consider the simplest polynomial case $V[X]=\frac{X^3}{3!}$, where $V'''[X]\neq0$.
The Hamiltonian density becomes
\begin{eqnarray}&&\label{CubicX}
\mathcal{H}_V=\frac{X^2}{3!}(X+6B_0^2)=\frac{X^2}{3!}(\vec{B}^2+5B_0^2+s b^2).
\end{eqnarray}
The candidate vacuum configuration $X=0$ forms a three-dimensional hyperbola. 
The flat direction corresponds to the tangent vector on this hypersurface $X=0$, satisfying $\bar{B}_\mu\delta B^\mu=0$, {\it, i.e.}, Eq. (\ref{FlatS}).
In fact, Eq. (\ref{CubicX}) is a special case of Eq. (\ref{HVgeneral}) in the main text,
and for $s=0,+1$, one can verify that $X=0$ is a global minimum.

\section{On-shell Hamiltonian of $p$-form fields with potential}
We begin the derivation from the action \eq{eq:p_form_action}.
The Lagrangian density of the $p$-form field is
\begin{align*}
\mathcal{L}&= -\frac{1}{2(p+1)!} F_{\mu_1 \dots \mu_{p+1}} F^{\mu_1 \dots \mu_{p+1}} - V(X)\\
&=\frac{1}{2 p!} (F_{0 i_1 \dots i_{p}})^2-\frac{1}{2(p+1)!} (F_{i_1 \dots i_{p+1}})^2 - V(X),
\end{align*}
where $i_1, \dots, i_{p+1}$ are spatial indices.
From $F_{\mu_1 \mu_2 \dots \mu_{p+1}} = (p+1) \partial_{[\mu_1} A_{\mu_2 \dots \mu_{p+1}]}$, we have
\begin{equation*}
F_{0 i_1\dots i_p}=\dot{A}_{i_1\dots i_p}-p\partial_{[ i_1}A_{|0|i_2\dots i_p]},
\end{equation*}
So the corresponding conjugate momenta are given as
\begin{equation*}
\Pi_{i_1 \dots i_p} = \frac{\partial \mathcal{L}}{\partial \dot{A}_{i_1 \dots i_p}} = \frac{1}{p!} F_{0 i_1 \dots i_p},
\end{equation*}
and we have the relation
\begin{equation*}
\dot{A}_{i_1\dots i_p}=F_{0 i_1\dots i_p}+p\partial_{[ i_1}A_{|0|i_2\dots i_p]}=p! \Pi_{i_1 \dots i_p}+p\partial_{[ i_1}A_{|0|i_2\dots i_p]}.
\end{equation*}
The Hamiltonian density is given as
\begin{align*}
\mathcal{H}&=
\Pi_{i_1 \dots i_p} \dot{A}_{i_1\dots i_p}-\mathcal{L}
\\&
=\Pi_{i_1 \dots i_p}(p! \Pi_{i_1 \dots i_p}+p\partial_{[ i_1}A_{|0|i_2\dots i_p]}) 
-\frac{1}{2 p!} (p! \Pi_{i_1 \dots i_p})^2+\frac{1}{2(p+1)!} (F_{i_1 \dots i_{p+1}})^2 + V(X)
\\&
=\frac{p!}{2}(\Pi_{i_1 \dots i_p})^2+\frac{1}{2(p+1)!} (F_{i_1 \dots i_{p+1}})^2 + V(X)+p \Pi_{i_1 \dots i_p}\partial_{[ i_1}A_{|0|i_2\dots i_p]}
\\&
=\frac{p!}{2}(\Pi_{i_1 \dots i_p})^2+\frac{1}{2(p+1)!} (F_{i_1 \dots i_{p+1}})^2 + V(X)+p \Pi_{i_1 \dots i_p}\partial_{ i_1}A_{0i_2\dots i_p}.
\end{align*}
After integrating by parts, we obtain
\begin{equation*}
\mathcal{H}=\frac{p!}{2}(\Pi_{i_1 \dots i_p})^2+\frac{1}{2(p+1)!} (F_{i_1 \dots i_{p+1}})^2 + V(X)-p \partial_{ i_1}\Pi_{i_1 i_2 \dots i_p}A_{0i_2\dots i_p}.
\end{equation*}

Now we apply the on-shell condition.
The equation of motion of the action \eq{eq:p_form_action} is
\begin{equation*}
\partial^{\nu} F_{\nu \mu_1 \dots \mu_p}=2p! V'(X) A_{\mu_1\dots \mu_p}.
\end{equation*}
Since $\nu, \mu_1, \dots, \mu_p$ are distinct, setting $\mu_1 = 0$, we have
\begin{equation*}
\partial_{i_1} F_{i_1 0 i_2 \dots i_p}=2p! V'(X) A_{0 i_2\dots i_p},
\end{equation*}
hence we have
\begin{equation*}
\partial_{ i_1}\Pi_{i_1 i_2 \dots i_p}=\frac{1}{p!} \partial_{ i_1}F_{0i_1 i_2 \dots i_p}=-\frac{1}{p!}  \partial_{ i_1}F_{i_10 i_2 \dots i_p}=-2V'(X) A_{0 i_2\dots i_p}.
\end{equation*}
Finally we obtain the on-shell Hamiltonian density as
\begin{equation*}
\mathcal{H}=\frac{p!}{2}(\Pi_{i_1 \dots i_p})^2+\frac{1}{2(p+1)!} (F_{i_1 \dots i_{p+1}})^2 + V(X)+2p (A_{0i_2\dots i_p})^2.
\end{equation*}

\bibliography{refs}

\begin{thebibliography}{18}%
\makeatletter
\providecommand \@ifxundefined [1]{%
 \@ifx{#1\undefined}
}%
\providecommand \@ifnum [1]{%
 \ifnum #1\expandafter \@firstoftwo
 \else \expandafter \@secondoftwo
 \fi
}%
\providecommand \@ifx [1]{%
 \ifx #1\expandafter \@firstoftwo
 \else \expandafter \@secondoftwo
 \fi
}%
\providecommand \natexlab [1]{#1}%
\providecommand \enquote  [1]{``#1''}%
\providecommand \bibnamefont  [1]{#1}%
\providecommand \bibfnamefont [1]{#1}%
\providecommand \citenamefont [1]{#1}%
\providecommand \href@noop [0]{\@secondoftwo}%
\providecommand \href [0]{\begingroup \@sanitize@url \@href}%
\providecommand \@href[1]{\@@startlink{#1}\@@href}%
\providecommand \@@href[1]{\endgroup#1\@@endlink}%
\providecommand \@sanitize@url [0]{\catcode `\\12\catcode `\$12\catcode
  `\&12\catcode `\#12\catcode `\^12\catcode `\_12\catcode `\%12\relax}%
\providecommand \@@startlink[1]{}%
\providecommand \@@endlink[0]{}%
\providecommand \url  [0]{\begingroup\@sanitize@url \@url }%
\providecommand \@url [1]{\endgroup\@href {#1}{\urlprefix }}%
\providecommand \urlprefix  [0]{URL }%
\providecommand \Eprint [0]{\href }%
\providecommand \doibase [0]{https://doi.org/}%
\providecommand \selectlanguage [0]{\@gobble}%
\providecommand \bibinfo  [0]{\@secondoftwo}%
\providecommand \bibfield  [0]{\@secondoftwo}%
\providecommand \translation [1]{[#1]}%
\providecommand \BibitemOpen [0]{}%
\providecommand \bibitemStop [0]{}%
\providecommand \bibitemNoStop [0]{.\EOS\space}%
\providecommand \EOS [0]{\spacefactor3000\relax}%
\providecommand \BibitemShut  [1]{\csname bibitem#1\endcsname}%
\let\auto@bib@innerbib\@empty
\bibitem [{\citenamefont {Kosteleck\'y}\ and\ \citenamefont
  {Samuel}(1989{\natexlab{a}})}]{SBLV1989a}%
  \BibitemOpen
  \bibfield  {author} {\bibinfo {author} {\bibfnamefont {V.~A.}\ \bibnamefont
  {Kosteleck\'y}}\ and\ \bibinfo {author} {\bibfnamefont {S.}~\bibnamefont
  {Samuel}},\ }\bibfield  {title} {\bibinfo {title} {Spontaneous breaking of
  lorentz symmetry in string theory},\ }\href
  {https://doi.org/10.1103/PhysRevD.39.683} {\bibfield  {journal} {\bibinfo
  {journal} {Phys. Rev. D}\ }\textbf {\bibinfo {volume} {39}},\ \bibinfo
  {pages} {683} (\bibinfo {year} {1989}{\natexlab{a}})}\BibitemShut {NoStop}%
\bibitem [{\citenamefont {Kosteleck\'y}\ and\ \citenamefont
  {Samuel}(1989{\natexlab{b}})}]{SBLV1989b}%
  \BibitemOpen
  \bibfield  {author} {\bibinfo {author} {\bibfnamefont {V.~A.}\ \bibnamefont
  {Kosteleck\'y}}\ and\ \bibinfo {author} {\bibfnamefont {S.}~\bibnamefont
  {Samuel}},\ }\bibfield  {title} {\bibinfo {title} {Gravitational
  phenomenology in higher-dimensional theories and strings},\ }\href
  {https://doi.org/10.1103/PhysRevD.40.1886} {\bibfield  {journal} {\bibinfo
  {journal} {Phys. Rev. D}\ }\textbf {\bibinfo {volume} {40}},\ \bibinfo
  {pages} {1886} (\bibinfo {year} {1989}{\natexlab{b}})}\BibitemShut {NoStop}%
\bibitem [{\citenamefont {Colladay}\ and\ \citenamefont
  {Kostelecky}(1998)}]{Colladay:1998fq}%
  \BibitemOpen
  \bibfield  {author} {\bibinfo {author} {\bibfnamefont {D.}~\bibnamefont
  {Colladay}}\ and\ \bibinfo {author} {\bibfnamefont {V.~A.}\ \bibnamefont
  {Kostelecky}},\ }\bibfield  {title} {\bibinfo {title} {{Lorentz violating
  extension of the standard model}},\ }\href
  {https://doi.org/10.1103/PhysRevD.58.116002} {\bibfield  {journal} {\bibinfo
  {journal} {Phys. Rev. D}\ }\textbf {\bibinfo {volume} {58}},\ \bibinfo
  {pages} {116002} (\bibinfo {year} {1998})},\ \Eprint
  {https://arxiv.org/abs/hep-ph/9809521} {arXiv:hep-ph/9809521} \BibitemShut
  {NoStop}%
\bibitem [{\citenamefont {Kosteleck\'y}(2004)}]{Gravity04}%
  \BibitemOpen
  \bibfield  {author} {\bibinfo {author} {\bibfnamefont {V.~A.}\ \bibnamefont
  {Kosteleck\'y}},\ }\bibfield  {title} {\bibinfo {title} {Gravity, lorentz
  violation, and the standard model},\ }\href
  {https://doi.org/10.1103/PhysRevD.69.105009} {\bibfield  {journal} {\bibinfo
  {journal} {Phys. Rev. D}\ }\textbf {\bibinfo {volume} {69}},\ \bibinfo
  {pages} {105009} (\bibinfo {year} {2004})}\BibitemShut {NoStop}%
\bibitem [{\citenamefont {Bluhm}\ and\ \citenamefont
  {Kosteleck\'y}(2005)}]{Bumblebee05}%
  \BibitemOpen
  \bibfield  {author} {\bibinfo {author} {\bibfnamefont {R.}~\bibnamefont
  {Bluhm}}\ and\ \bibinfo {author} {\bibfnamefont {V.~A.}\ \bibnamefont
  {Kosteleck\'y}},\ }\bibfield  {title} {\bibinfo {title} {Spontaneous lorentz
  violation, nambu-goldstone modes, and gravity},\ }\href
  {https://doi.org/10.1103/PhysRevD.71.065008} {\bibfield  {journal} {\bibinfo
  {journal} {Phys. Rev. D}\ }\textbf {\bibinfo {volume} {71}},\ \bibinfo
  {pages} {065008} (\bibinfo {year} {2005})}\BibitemShut {NoStop}%
\bibitem [{\citenamefont {Bluhm}\ \emph
  {et~al.}(2008{\natexlab{a}})\citenamefont {Bluhm}, \citenamefont {Gagne},
  \citenamefont {Potting},\ and\ \citenamefont {Vrublevskis}}]{Bumblebee08}%
  \BibitemOpen
  \bibfield  {author} {\bibinfo {author} {\bibfnamefont {R.}~\bibnamefont
  {Bluhm}}, \bibinfo {author} {\bibfnamefont {N.~L.}\ \bibnamefont {Gagne}},
  \bibinfo {author} {\bibfnamefont {R.}~\bibnamefont {Potting}},\ and\ \bibinfo
  {author} {\bibfnamefont {A.}~\bibnamefont {Vrublevskis}},\ }\bibfield
  {title} {\bibinfo {title} {Constraints and stability in vector theories with
  spontaneous lorentz violation},\ }\href
  {https://doi.org/10.1103/PhysRevD.77.125007} {\bibfield  {journal} {\bibinfo
  {journal} {Phys. Rev. D}\ }\textbf {\bibinfo {volume} {77}},\ \bibinfo
  {pages} {125007} (\bibinfo {year} {2008}{\natexlab{a}})}\BibitemShut
  {NoStop}%
\bibitem [{\citenamefont {Mattingly}(2005)}]{Mattingly:2005re}%
  \BibitemOpen
  \bibfield  {author} {\bibinfo {author} {\bibfnamefont {D.}~\bibnamefont
  {Mattingly}},\ }\bibfield  {title} {\bibinfo {title} {{Modern tests of
  Lorentz invariance}},\ }\href {https://doi.org/10.12942/lrr-2005-5}
  {\bibfield  {journal} {\bibinfo  {journal} {Living Rev. Rel.}\ }\textbf
  {\bibinfo {volume} {8}},\ \bibinfo {pages} {5} (\bibinfo {year} {2005})},\
  \Eprint {https://arxiv.org/abs/gr-qc/0502097} {arXiv:gr-qc/0502097}
  \BibitemShut {NoStop}%
\bibitem [{\citenamefont {Tasson}(2014)}]{Tasson:2014dfa}%
  \BibitemOpen
  \bibfield  {author} {\bibinfo {author} {\bibfnamefont {J.~D.}\ \bibnamefont
  {Tasson}},\ }\bibfield  {title} {\bibinfo {title} {{What Do We Know About
  Lorentz Invariance?}},\ }\href
  {https://doi.org/10.1088/0034-4885/77/6/062901} {\bibfield  {journal}
  {\bibinfo  {journal} {Rept. Prog. Phys.}\ }\textbf {\bibinfo {volume} {77}},\
  \bibinfo {pages} {062901} (\bibinfo {year} {2014})},\ \Eprint
  {https://arxiv.org/abs/1403.7785} {arXiv:1403.7785 [hep-ph]} \BibitemShut
  {NoStop}%
\bibitem [{\citenamefont {Bluhm}\ \emph
  {et~al.}(2008{\natexlab{b}})\citenamefont {Bluhm}, \citenamefont {Gagne},
  \citenamefont {Potting},\ and\ \citenamefont {Vrublevskis}}]{Bluhm:2008yt}%
  \BibitemOpen
  \bibfield  {author} {\bibinfo {author} {\bibfnamefont {R.}~\bibnamefont
  {Bluhm}}, \bibinfo {author} {\bibfnamefont {N.~L.}\ \bibnamefont {Gagne}},
  \bibinfo {author} {\bibfnamefont {R.}~\bibnamefont {Potting}},\ and\ \bibinfo
  {author} {\bibfnamefont {A.}~\bibnamefont {Vrublevskis}},\ }\bibfield
  {title} {\bibinfo {title} {{Constraints and Stability in Vector Theories with
  Spontaneous Lorentz Violation}},\ }\href
  {https://doi.org/10.1103/PhysRevD.79.029902} {\bibfield  {journal} {\bibinfo
  {journal} {Phys. Rev. D}\ }\textbf {\bibinfo {volume} {77}},\ \bibinfo
  {pages} {125007} (\bibinfo {year} {2008}{\natexlab{b}})},\ \bibinfo {note}
  {[Erratum: Phys.Rev.D 79, 029902 (2009)]},\ \Eprint
  {https://arxiv.org/abs/0802.4071} {arXiv:0802.4071 [hep-th]} \BibitemShut
  {NoStop}%
\bibitem [{\citenamefont {Carroll}\ \emph {et~al.}(2009)\citenamefont
  {Carroll}, \citenamefont {Dulaney}, \citenamefont {Gresham},\ and\
  \citenamefont {Tam}}]{Carroll:2009em}%
  \BibitemOpen
  \bibfield  {author} {\bibinfo {author} {\bibfnamefont {S.~M.}\ \bibnamefont
  {Carroll}}, \bibinfo {author} {\bibfnamefont {T.~R.}\ \bibnamefont
  {Dulaney}}, \bibinfo {author} {\bibfnamefont {M.~I.}\ \bibnamefont
  {Gresham}},\ and\ \bibinfo {author} {\bibfnamefont {H.}~\bibnamefont {Tam}},\
  }\bibfield  {title} {\bibinfo {title} {{Instabilities in the Aether}},\
  }\href {https://doi.org/10.1103/PhysRevD.79.065011} {\bibfield  {journal}
  {\bibinfo  {journal} {Phys. Rev. D}\ }\textbf {\bibinfo {volume} {79}},\
  \bibinfo {pages} {065011} (\bibinfo {year} {2009})},\ \Eprint
  {https://arxiv.org/abs/0812.1049} {arXiv:0812.1049 [hep-th]} \BibitemShut
  {NoStop}%
\bibitem [{\citenamefont {Bailey}\ \emph {et~al.}(2025)\citenamefont {Bailey},
  \citenamefont {Murray},\ and\ \citenamefont
  {Walter-Cardona}}]{Bailey:2025oun}%
  \BibitemOpen
  \bibfield  {author} {\bibinfo {author} {\bibfnamefont {Q.~G.}\ \bibnamefont
  {Bailey}}, \bibinfo {author} {\bibfnamefont {H.~S.}\ \bibnamefont {Murray}},\
  and\ \bibinfo {author} {\bibfnamefont {D.~T.}\ \bibnamefont
  {Walter-Cardona}},\ }\bibfield  {title} {\bibinfo {title} {{Bumblebee
  gravity: Spherically symmetric solutions away from the potential minimum}},\
  }\href {https://doi.org/10.1103/nkbj-vbjf} {\bibfield  {journal} {\bibinfo
  {journal} {Phys. Rev. D}\ }\textbf {\bibinfo {volume} {112}},\ \bibinfo
  {pages} {024069} (\bibinfo {year} {2025})},\ \Eprint
  {https://arxiv.org/abs/2503.10998} {arXiv:2503.10998 [gr-qc]} \BibitemShut
  {NoStop}%
\bibitem [{\citenamefont {Kraus}\ and\ \citenamefont
  {Tomboulis}(2002)}]{SLVScenario02}%
  \BibitemOpen
  \bibfield  {author} {\bibinfo {author} {\bibfnamefont {P.}~\bibnamefont
  {Kraus}}\ and\ \bibinfo {author} {\bibfnamefont {E.~T.}\ \bibnamefont
  {Tomboulis}},\ }\bibfield  {title} {\bibinfo {title} {Photons and gravitons
  as goldstone bosons and the cosmological constant},\ }\href
  {https://doi.org/10.1103/PhysRevD.66.045015} {\bibfield  {journal} {\bibinfo
  {journal} {Phys. Rev. D}\ }\textbf {\bibinfo {volume} {66}},\ \bibinfo
  {pages} {045015} (\bibinfo {year} {2002})}\BibitemShut {NoStop}%
\bibitem [{\citenamefont {Englert}\ and\ \citenamefont
  {Brout}(1964)}]{Englert:1964et}%
  \BibitemOpen
  \bibfield  {author} {\bibinfo {author} {\bibfnamefont {F.}~\bibnamefont
  {Englert}}\ and\ \bibinfo {author} {\bibfnamefont {R.}~\bibnamefont
  {Brout}},\ }\bibfield  {title} {\bibinfo {title} {{Broken Symmetry and the
  Mass of Gauge Vector Mesons}},\ }\href
  {https://doi.org/10.1103/PhysRevLett.13.321} {\bibfield  {journal} {\bibinfo
  {journal} {Phys. Rev. Lett.}\ }\textbf {\bibinfo {volume} {13}},\ \bibinfo
  {pages} {321} (\bibinfo {year} {1964})}\BibitemShut {NoStop}%
\bibitem [{\citenamefont {Goldstone}(1961)}]{Goldstone:1961eq}%
  \BibitemOpen
  \bibfield  {author} {\bibinfo {author} {\bibfnamefont {J.}~\bibnamefont
  {Goldstone}},\ }\bibfield  {title} {\bibinfo {title} {{Field Theories with
  Superconductor Solutions}},\ }\href {https://doi.org/10.1007/BF02812722}
  {\bibfield  {journal} {\bibinfo  {journal} {Nuovo Cim.}\ }\textbf {\bibinfo
  {volume} {19}},\ \bibinfo {pages} {154} (\bibinfo {year} {1961})}\BibitemShut
  {NoStop}%
\bibitem [{\citenamefont {Weinberg}(1995)}]{Weinberg:1996kr}%
  \BibitemOpen
  \bibfield  {author} {\bibinfo {author} {\bibfnamefont {S.}~\bibnamefont
  {Weinberg}},\ }\href@noop {} {\emph {\bibinfo {title} {{The Quantum Theory of
  Fields. Vol. 2: Modern Applications}}}}\ (\bibinfo  {publisher} {Cambridge
  University Press},\ \bibinfo {year} {1995})\ \bibinfo {note} {chapter
  19}\BibitemShut {NoStop}%
\bibitem [{\citenamefont {Jackiw}(1974)}]{Jackiw:1974cv}%
  \BibitemOpen
  \bibfield  {author} {\bibinfo {author} {\bibfnamefont {R.}~\bibnamefont
  {Jackiw}},\ }\bibfield  {title} {\bibinfo {title} {{Functional evaluation of
  the effective potential}},\ }\href {https://doi.org/10.1103/PhysRevD.9.1686}
  {\bibfield  {journal} {\bibinfo  {journal} {Phys. Rev. D}\ }\textbf {\bibinfo
  {volume} {9}},\ \bibinfo {pages} {1686} (\bibinfo {year} {1974})}\BibitemShut
  {NoStop}%
\bibitem [{\citenamefont {Dirac}(1950)}]{Dirac:1950pj}%
  \BibitemOpen
  \bibfield  {author} {\bibinfo {author} {\bibfnamefont {P.~A.~M.}\
  \bibnamefont {Dirac}},\ }\bibfield  {title} {\bibinfo {title} {{Generalized
  Hamiltonian dynamics}},\ }\href {https://doi.org/10.4153/CJM-1950-012-1}
  {\bibfield  {journal} {\bibinfo  {journal} {Can. J. Math.}\ }\textbf
  {\bibinfo {volume} {2}},\ \bibinfo {pages} {129} (\bibinfo {year}
  {1950})}\BibitemShut {NoStop}%
\bibitem [{\citenamefont {Bonder}\ and\ \citenamefont
  {Escobar}(2016)}]{YBonder2015E}%
  \BibitemOpen
  \bibfield  {author} {\bibinfo {author} {\bibfnamefont {Y.}~\bibnamefont
  {Bonder}}\ and\ \bibinfo {author} {\bibfnamefont {C.~A.}\ \bibnamefont
  {Escobar}},\ }\bibfield  {title} {\bibinfo {title} {{Dynamical ambiguities in
  models with spontaneous Lorentz violation}},\ }\href
  {https://doi.org/10.1103/PhysRevD.93.025020} {\bibfield  {journal} {\bibinfo
  {journal} {Phys. Rev. D}\ }\textbf {\bibinfo {volume} {93}},\ \bibinfo
  {pages} {025020} (\bibinfo {year} {2016})},\ \Eprint
  {https://arxiv.org/abs/1510.05999} {arXiv:1510.05999 [hep-th]} \BibitemShut
  {NoStop}%
\end{thebibliography}%

\end{document}